# Minimizing Fleet Size and Improving Bike Allocation of Bike Sharing under Future Uncertainty

**Mingzhuang Hua, Ph.D. candidate**
School of Transportation, Southeast University
Department of Technology, Management and Economics, Technical University of Denmark
Bygningstorvet Building 116, Kgs. Lyngby, Denmark, 2800
Email: huamingzhuang@seu.edu.cn

**Xuewu Chen, Ph.D., Professor (Corresponding Author)**
Jiangsu Key Laboratory of Urban ITS, Southeast University
Jiangsu Province Collaborative Innovation Center of Modern Urban Traffic Technologies, Southeast University
School of Transportation, Southeast University
Dongnandaxue Road #2, Nanjing, China, 211189
Email: chenxuewu@seu.edu.cn

**Jingxu Chen, Ph.D., Associate professor**
School of Transportation, Southeast University
Dongnandaxue Road #2, Nanjing, China, 211189
Email: chenjingxu@seu.edu.cn

**Yu Jiang, Ph.D., Associate professor**
Department of Technology, Management and Economics, Technical University of Denmark
Bygningstorvet Building 116, Kgs. Lyngby, Denmark, 2800
Email: yujiang@dtu.dk

# Minimizing Fleet Size and Improving Bike Allocation of Bike Sharing under Future Uncertainty

**ABSTRACT**

As a rapidly expanding service, bike sharing is facing severe problems of bike over-supply and demand fluctuation in many Chinese cities. This study develops a large-scale method to determine the minimum fleet size under uncertainty, based on the bike sharing data of millions of trips in Nanjing. It is found that the algorithm of minimizing fleet size under the incomplete-information scenario is effective in handling future uncertainty. For a dockless bike sharing system, supplying 14.5% of the original fleet could meet 96.8% of trip demands. Meanwhile, the results suggest that providing a integrated service platform that integrates multiple companies can significantly reduce the total fleet size by 44.6%. Moreover, in view of the COVID-19 pandemic, this study proposes a social distancing policy that maintains a suitable usage interval. These findings provide useful insights for improving the resource efficiency and operational service of bike sharing and shared mobility.

# 1. Introduction

Bike sharing has been under rapid development for the last few years, as it is a successful application of the emerging shared mobility, and it has a great potential to alleviate traffic congestion (Wang and Zhou, 2017), promote physical health (Otero et al., 2018), reduce greenhouse gas emissions (Luo et al., 2019; Kou et al., 2020; Chen et al., 2020), and improve resource efficiency (Böckin et al., 2020). There are two types of bike-sharing systems: station-based bike sharing (SBBS) and dockless bike sharing (DBS). With the advances in cashless mobile payments and location tracking technology, DBS has experienced a period of rapid expansion (WIRED, 2017) and fierce competition (Choudhury, 2018).

One of the key challenges of operators' desire for a high market share is the bike over-supply issue (Taylor, 2018) of the large-scale fleet. This has even created bike graveyards in severe cases (Rong et al., 2019; Chen et al., 2020), leading to an increase in the operation cost. Hence, it is necessary for operators to avoid idle shared bikes for maintaining operational efficiency. Meanwhile, it is also vital that the fleet size should cater to the demands of bike sharing to retain existing users and attract new users. In general, determining such a proper fleet size requires answering the following two questions. How many shared bikes does a city need at least to meet user demands? Where should these bikes be allocated?

Another key challenge in the bike sharing operation is that the demands of the bike sharing system are random and unpredictable (Jie et al., 2020). This study focuses on future uncertainty, which includes two aspects. One is the variation in the quantity and the other is the uncertainty of the origin and destination locations. The former variation applies to both SBBS and DBS, while the latter one is particularly related to the DBS. To address this problem, this study develops a novel methodology for determining the least number of bikes under the future uncertainty.

Besides, bike sharing emerges as an important mobility option after the COVID-19 outbreak (Teixeira and Lopes, 2020; Bucsky, 2020), even though this pandemic has a dramatic impact on urban transportation. Meeting travel demand and ensuring safety are difficult trade-offs for bike sharing services during the COVID-19 (Hua et al., 2021). Bike allocation is the key challenge to this problem, especially to reduce the risk of virus transmission from user exposure to bikes. Given that coronavirus stability decays over time, avoiding different users using the same bike in a short period is one possible solution.

Vehicle allocation (espically fleet size) is the core problem of shared mobility, which lays the foundation for facility planning and operation management. Specially, minimizing fleet size is a pressing problem for bike-sharing, given the severe issue of bike oversupply. Existing fleet size studies have focused on the small-scale problem or relied on the unrealistic assumption, resulting in their ineffective application to company operation, as described in Section 2. Our approach addresses the large-scale problem and deals with future uncertainty.

This paper is organized as follows. Section 2 introduces the existing studies related to resource allocation. Section 3 elaborates the bike allocation method for minimizing fleet size and improving bike allocation. Section 4 takes the Nanjing bike sharing (with and without docks) as an example to show the application results of our allocation method. Section 5 conducts the sensitivity analysis of user distancing and a single platform. Finally, Section 6 concludes this study and points out future research directions.

## 2. Literature review

### 2.1. Resource allocation of bike sharing

Resource allocation studies of bike sharing include station planning and bike rebalancing, except for fleet size. Station planning studies mainly apply optimization models and geographic information system(GIS)-based methods. Yuan et al. (2019) proposed a mixed integer linear programming (MILP) model for station planning and applied it in a small area to select 32 stations from 100 grids. Loidl et al. (2019) established a GIS method based on spatial data and suggested 100 stations in a city case. Hu et al. (2019) used the capacitated maximal covering location problem to decide the optimal bike station location and applied it in the case of 40 stations. Caggiani et al. (2020) built a bike station location model for multimodal bike-public transport and used a genetic algorithm to select 30 stations from 200 potential sites in the numerical case. Eren and Katanalp (2022) designed a GIS-based multi-criteria decision-making method to select stations and chose the best 5 stations from 42 potential locations in the case study. Fu et al. (2022) proposed a customized row generation algorithm to solve the station location problem considering rebalancing and applied it in the numerical examples of 40 stations. In short, existing station planning studies rarely address thousands of bike stations because of lacking station planning methods for processing big data with millions of trips.

Many existing studies of bike rebalancing have established various optimization models and solution methods. The rebalancing algorithms can be solved via both exact and heuristic algorithms (Kroes et al., 2020), as the bike rebalancing problem is NP-hard (Lv et al., 2020). Exact algorithms are suitable for a small-scale problem, while heuristic algorithms could solve a large-scale problem for practical application. Bike rebalancing includes static rebalancing (Du et al., 2020; Yaping Ren et al., 2020; Ma et al., 2021) and dynamic rebalancing (Legros, 2019; Hu et al., 2021; He et al., 2021). Static rebalancing occurs during late night hours, while dynamic rebalancing occurs during the daytime. Bruck et al. (2019) proposed an exact algorithm based on a minimal extended network to solve the static rebalancing problem and applied it in the instances of 80 stations. Tian et al. (2020) designed a flow-type task window method for the dynamic rebalancing problem. Their case study shows that task window and region division could reduce the station amount to 19 and therefore simplify the rebalancing problem. While dynamic rebalancing can respond to user demands in a timely manner, static rebalancing is also a core aspect of operation management. The difference between static rebalancing and dynamic rebalancing lies not in the importance for operation management but in different operation periods (midnight vs. daytime). Static rebalancing determines the initial inventory level each day before users start riding, which decides the base of meeting travel demand and performing dynamic rebalancing. The result of static rebalancing is the state of fleet allocation, so static rebalancing is the important part of this fleet size problem.

## 2.2. Fleet size of shared mobility

Existing studies of shared mobility on fleet size mainly focus on economic theory and mathematical programming. Yang et al. (2014) employed the Cobb-Douglas meeting function to estimate the returns to scale in taxi services and proved that increasing taxi fleet size would decrease waiting time and increase customer demand. Masson et al. (2017) proposed a mathematical model to optimize vehicle amounts for the mixed transportation system with passengers and goods and apply a meta-heuristic method called adaptive large neighborhood search in a bus line case with 6 stations. Xu et al. (2018) proposed a mixed-integer nonlinear nonconvex model to optimize the fleet size and trip pricing for car sharing service. They found that providing more stations can meet more demands and increase profits in the Singapore case. Simonetto et al. (2019) developed a federated optimization architecture to solve the linear assignment problem of ride sourcing service and discovered that the negative competition of multiple companies would increase the fleet size. Dlugosch et al. (2020) estimated the impact of shared autonomous electric vehicles and found that autonomous vehicles can reduce about half of fleet size in carsharing. Nevertheless, economic theory cannot determine the precise spatial distribution of shared mobility, and mathematical programming is not suitable for solving the large-scale problem for resource allocation with millions of trips and thousands of stations.

Existing research of bike sharing on fleet size (also known as inventory level or bike allocation) focuses on station-based bike sharing and small-scale cases under one thousand stations. So their methods have not adequately considered practical feasibility in city-wide bike sharing management. Gómez Márquez et al. (2021) proposed the simulation-optimization heuristic algorithms to optimize initial bike inventory and tested their method in the case study of 500 stations. But their method would reject about 15% of trips, and this unmet trip ratio is too high to be practicable in the company operation. Soriguera and Jiménez-Meroño (2020) estimated station amount and fleet size with a continuous approximation model, which has been validated in the bike sharing dataset of 402 stations. Datner et al. (2019) proposed a robust guided local search method for setting target inventory levels and applied this algorithm in a numerical study of about 300 stations. Caggiani et al. (2019) proposed an optimization model of maximizing user satisfaction to decide the dock and bike allocation and applied it in the numerical example of 144 stations. Ren et al. (2020) established a MILP model and a branch-and-cut algorithm for minimizing the bike amount in a depot, and their method is applied in a real-world instance of 13 stations. Swaszek and Cassandras (2020) used the index of demand rate and station capacity to decide the inventory level, and tested their method by a case study of 11 stations. Maleki Vishkaei et al. (2020) used critical inventory levels by the closed queuing network to determine the dock and bike allocation and applied their approach in a case study of 4 stations. In actual company operation of bike sharing, bike allocation in a city should consider thousands of places (stations or other nodes) and the computational complexity (Pal and Zhang, 2017) of the free-floating DBS systems. Therefore, the large-scale problem and free-floating challenge are necessary for the city-scale bike allocation, which are not well solved in the existing studies.

*2.3. Minimum fleet size problem*

It is important to emphasize that minimizing fleet size and rebalancing are two different problems. Due to the market competition (Zhang and Meng, 2019) of bike sharing companies, the issue of bike over-supply (Tao and Zhou, 2021) has been caused. This issue has seriously exacerbated the imbalance of supply and demand and increased many unnecessary operation tasks. Rebalancing is only to solve the imbalance of supply and demand for the existing fleet. But minimizing fleet size can not only solves the issue of bike over-supply, but also correspondingly reduce rebalancing tasks greatly. If bike over-supply is like a disease, rebalancing can only relieve symptoms, while minimizing fleet size can cure the disease.

Even though many papers focus on the fleet size of shared mobility, only very few studies discuss minimizing the fleet size for large-scale problems. Vazifeh et al. (2018) provided the first solution for the minimum fleet size problem of city-scale shared mobility. They used Hopcroft–Karp algorithm to minimize the taxi fleet and applied the network-based solution in New York. With the completely accurate forecast of daily trip demands, their optimal solution can reduce taxi fleet size by 40%. Gu et al. (2019) proposed a heuristic algorithm to determine the optimal bike supply of dockless bike sharing (DBS) and tested this method on the DBS dataset in Shenzhen for two days. They found that less than 20% of the actual bike fleet can cover all the trips in these two days. Luo et al. (2020) established an integrated framework to determine the optimal fleet size and bike rebalancing strategy to minimize life cycle emissions and found that only 15% of the bike fleet could serve the same demands. However, these studies all assumed that the demands in the selected day are representative and will be repeated in the future, which means that operators could have the full knowledge of future trip demands. The complete-information assumption of future trip demands is unrealistic and has a huge negative impact on meeting user demands, which will be demonstrated and discussed in the subsequent parts of this study.

This paper simultaneously determines the minimum fleet size and the most reasonable bike allocation for meeting future trip demands and examines the method performance with the cases of the Nanjing bike sharing systems. To the best of our knowledge, this study is the first paper to solve the minimum bike fleet size problem with future uncertainty. The contributions of this study include: 1) Developing a general framework to solve the minimum fleet size problem under historical trip and future uncertainty; 2) Proposing to establish an integrated platform that is compatible with multiple bike sharing companies and examining its effect on the bike supply; 3) Proposing a bike allocation strategy according to the social distancing policy considering the pandemic impacts.

## 3. Methodology

In particular, there are three types of fleet size in bike sharing services, including actual fleet size, active fleet size, and minimum fleet size. (a) Actual fleet size is defined as the number of all bikes in actual operation, which is reported to the government by the companies. (b) Active fleet size calculates the number of bikes used at least once during a selected period. (c) Minimum fleet size is defined as the minimum bike amount when user demands are met, which is the fleet size to be determined in this study. Minimum fleet size is the core concept in operating bike sharing systems as it indicates that the number of bikes should be ready for rent at various places in this day. Thereby, it is necessary to take into both bike allocation and demand variation in the process of determining the minimum fleet size.

There are several basic components that need to be considered in determining the minimum fleet size for a large-scale bike sharing service. (a) Trip demand, which is the set of user travels from trip data, including origin, start time, destination, and end time. (b) Bike demand is defined as the number of bikes required at each place to meet all trip demands. Minimum fleet size is the sum of bike demands in all places each day. Generally, the trip demand is greater than the bike demand because one bike can serve the demands of multiple trips. Meanwhile, estimating the bike demands and minimizing the fleet size of DBS is more complicated than that of SBBS because DBS bikes are free-floating, and DBS trip demands are greater. (c) Bike rebalancing is defined as the operational activity of meeting the supply-demand gap, i.e., moving bicycles to where they are needed. The amount of bike rebalancing is equal to the sum of the move-in bike amounts at all places, and the move-in bicycle amount is the difference between the number of bike demands and the number of bikes available at a place. The linkages of the three components in operating bike sharing systems are shown in Fig. 1.

To determine the minimum fleet size, the input information in this study is the predicted trip data for the future period. In general, we use the input information to distinguish two scenarios: complete-information vs. incomplete-information. In the first scenario, it is assumed that we have complete information to predict the correct details of all trips for the next day. However, no one can predict the future with complete accuracy, so the complete-information assumption is impossible in actual operation. In the second scenario, we do not pose the complete-information for predicting correct future trips but only rely on the incomplete-information based on historical trip data. Correspondingly, we define two types of minimum fleet size, namely, *ideal* minimum fleet size and *approximated* minimum fleet size, associated with the two scenarios.

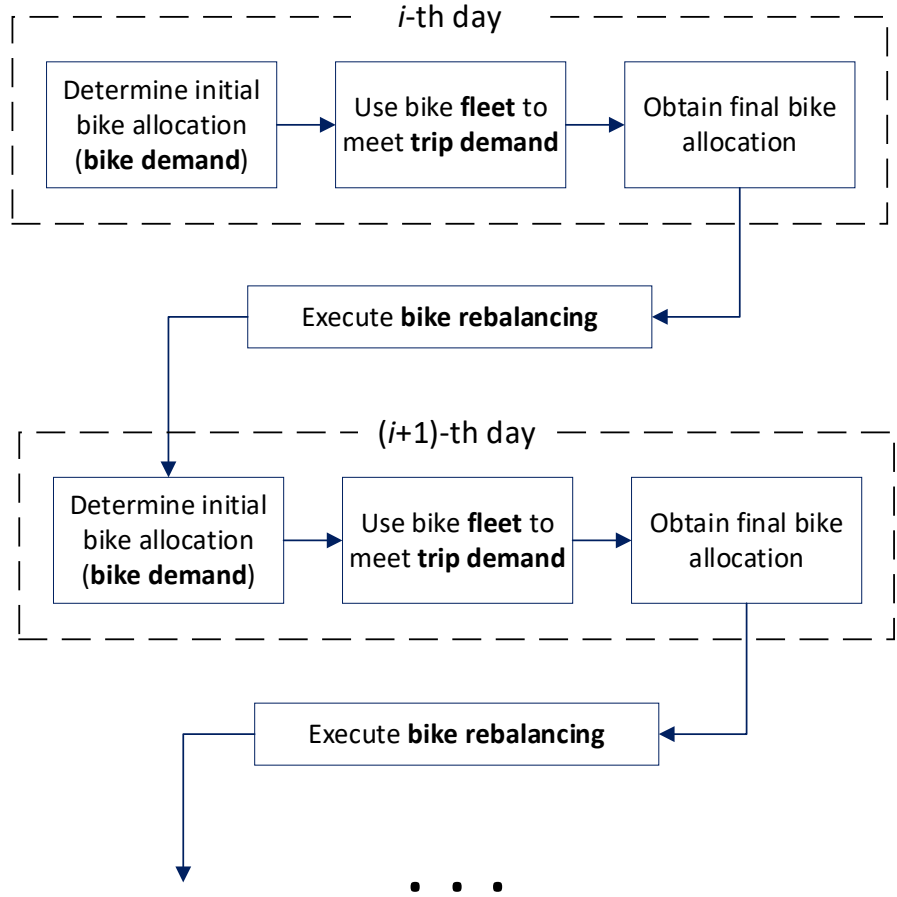

**Fig. 1.** The basic components in the bike allocation study.

On the one hand, the existing studies (Vazifeh et al., 2018; Gu et al., 2019; Luo et al., 2020) focus on the first scenario by assuming that the future demand is well known at the beginning of the day. The *ideal* minimum fleet size is under the complete-information assumption, which is the minimum bike amount when the future bike demands can be completely predicted and fully met. The algorithm for minimizing fleet size under the complete-information assumption is described in Section 3.1.

On the other hand, under the incomplete-information scenario, future demands cannot be exactly predicted. The operator relies on historical information to determine the bike allocation, so we use the historical trip data to estimate future demands. The *approximated* minimum fleet size is under an incomplete-information scenario, the minimum bike amount when the future bike demands can only be partially predicted and met as much as possible. In this study, the key to meeting uncertain demands is that the available historical trip data covers a longer period than the period to be predicted in the future. We evaluate the bike demands of each place in the past several days and select the maximum value as the allocated bike amount. This allows providing the necessary redundancy for meeting the demands on the following day, which could handle the uncertainty of future trip demand. To the best of our knowledge, such an issue has not been well addressed in existing studies, and our paper is the first one to tackle it. The algorithm for minimizing fleet size under an incomplete-information scenario is described in Section 3.2.

*3.1. Minimizing fleet size under the complete-information assumption*

The principle of minimizing fleet size is to assign all trips to as few bikes as possible. A reasonable solution is to develop an advanced greedy algorithm that considers spatio-temporal relationships to match trips and bikes. This greedy algorithm has three procedures. Firstly, add a bike to connect as many trips as possible according to time sequence and spatial proximity criterion. The spatial proximity criterion in SBBS is that the start position of the latter trip and the end position of the former trip are in the same station. And the spatial proximity criterion in DBS is that the distance from the start position of the latter trip to the end position of the former trip is less than the acceptable walking distance. Secondly, stop using this bike and delete these trips. Finally, repeat the first two steps until the trip dataset is empty. This greedy algorithm can obtain minimizing fleet algorithm under the complete-information assumption, and its steps are shown in Fig. 2.

Each trip in the trip dataset is represented by $T_i$ ($O_i$, $S_i$, $D_i$, $E_i$). $O_i$ is the origin (start station in SBBS, start longitude and start latitude in DBS) of $i$-th trip $T_i$. $D_i$ is the destination (end station in SBBS, end longitude and end latitude in DBS). $S_i$ is the start time, and $E_i$ is the end time. Each bike in the fleet is represented by $B_j$ ($A_j$, $C_j$, $P_j$, $H_j$). $A_j$ is the initial position of $j$-th bike $B_j$ on this day, and $C_j$ is the final position. $P_j$ is the position of this bike at time point $H_j$.

This study has other two parameters, including acceptable walk distance $w$ and bike usage interval $c$. Acceptable walking distance $w$ is the searching radius of the DBS bike position. If the distance between bike position and user origin is less than $w$, the user can walk to the bike location and use this bike for riding. Otherwise, the bike is too far away from the user to be used. In most previous studies (Fishman et al., 2015; Faghih-Imani and Eluru, 2016; Bao et al., 2017; Böcker et al., 2020; Kumar Dey et al., 2021), the searching radius of bike position is 250 meters. Fuller et al. (2011) conducted a bike sharing survey in Montreal, Canada, and found that using shared bikes or not depended on whether the respondent lived within 250 meters of the bike position. Thereby, the value of $w$ is set as 250 meters as well in this study

Bike usage interval $c$ is the minimum time interval between the former and latter users using the same bikes, which has different values in different operational states. In normal operation, shared bikes should avoid being idle in parking places. So the value of $c$ is set as 0 in this normal state. But during the COVID-19 pandemic, the former and latter users would touch the same handlebar and seat, which is a risky transmission pathway (Hua et al., 2021). A user distancing approach is proposed to maintain enough time intervals of these users, as the application of social distancing strategy in bike sharing. On these surfaces of stainless steel and plastic, the half-life of the coronavirus is about 6 hours (van Doremalen et al., 2020). So the value of $c$ is set as 6 hours as the user distancing strategy in the pandemic state, as shown in equation (1).

$$c = \begin{cases} 0, & \text{normal operation} \\ 6 \text{ hours}, & \text{user distancing} \end{cases} \quad (1)$$

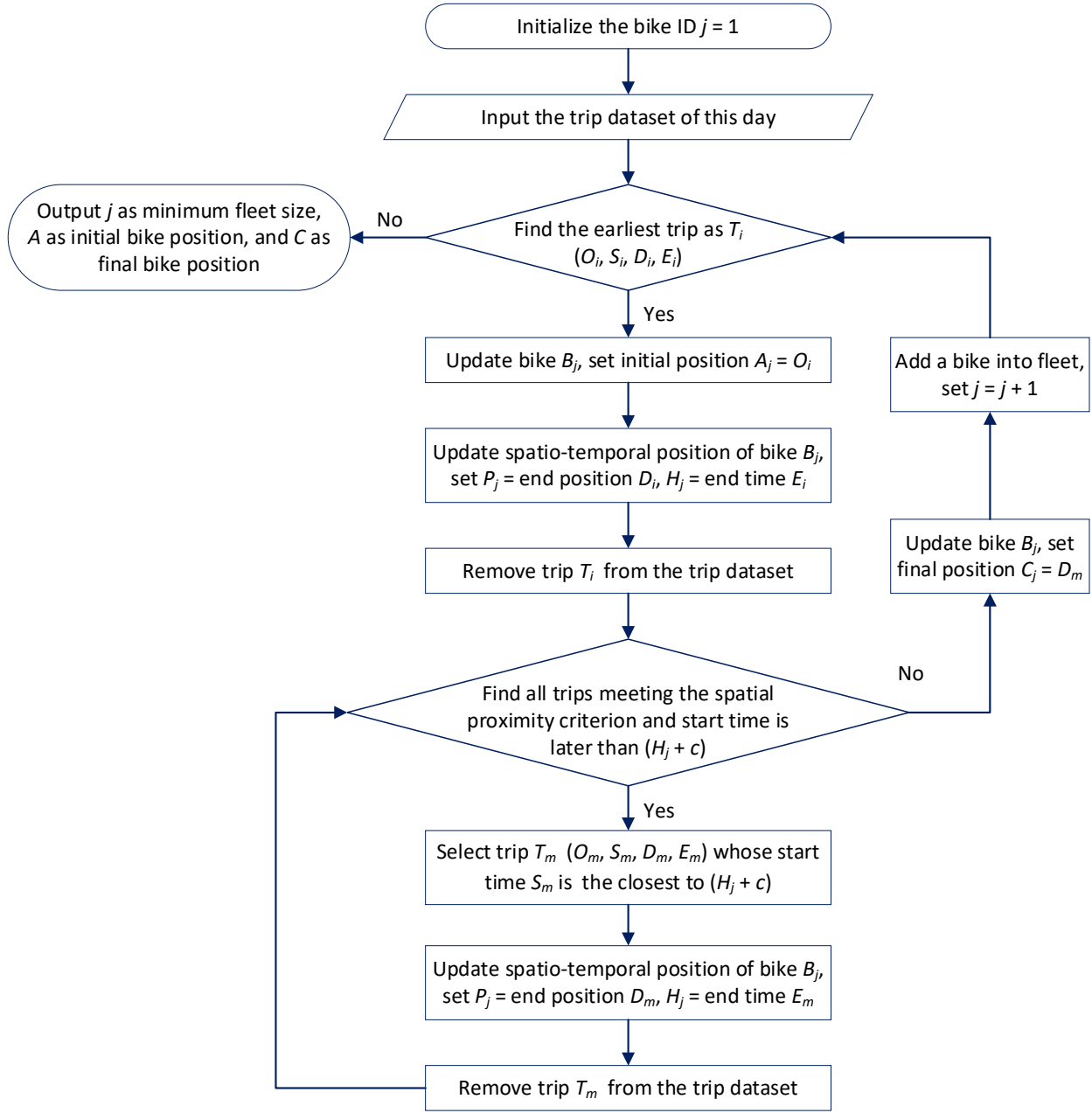

**Fig. 2.** Minimizing fleet size algorithm under the complete-information assumption.

For the bike allocation problem of SBBS, this greedy algorithm can get the minimum fleet size, which can be inferred as follows. An obvious finding is that in any place at any time point, breaking all trip connections and then arbitrarily reconnecting trips will not increase the bike amount. Firstly, assume that the minimum fleet size in bike-trip matching results has been obtained. Secondly, in any place at any time point, break all trip connections and then reconnect trips according to the greedy principle. Thirdly, repeat this break-reconnect step continuously. Finally, the bike-trip matching result of the greedy algorithm can be obtained. In the above conversion process, from the minimum fleet size to the greedy algorithm result, the bike amount does not increase. Therefore, the fleet size in this greedy algorithm is the same as the minimum fleet size. The detailed derivation process is shown in the Appendix.

For the bike allocation problem of DBS, this greedy algorithm may not get the minimum fleet size. In the SBBS service, bike positions are in the fixed stations, and the spatial proximity criterion is whether these trips are in the same station. So the break-reconnect step in SBBS will not increase the bike amount. But the DBS bike positions are free-floating, and the spatial proximity criterion of DBS is the search radius of 250 meters. Therefore, the break-reconnect step in DBS may increase the bike amount. However, this greedy algorithm will still significantly improve trip connection and bike turnover in DBS service, resulting in its fleet size being much smaller than the actual fleet size.

### 3.2. Minimizing fleet size under the incomplete-information scenario

No one is fully-informed of the future, so future trip demands and future bike demands cannot be completely predicted. An effective solution is calculating historical bike demands for estimating future bike demands. Therefore, there are two algorithms for minimizing fleet size: (a) minimizing fleet size under the complete-information assumption and (b) minimizing fleet size under the incomplete-information scenario. The algorithm of minimizing fleet size under the incomplete-information scenario is compatible with the algorithm of minimizing fleet size under the complete-information assumption. It means that minimizing fleet size under an incomplete-information scenario is a more general algorithm. This general framework of minimizing bike fleet size is shown in Fig. 3, and a key part of Fig. 3 is Fig. 2 above.

Minimizing fleet size under the incomplete-information scenario generally has three steps: (a) select proper spatial objects for fleet management of both SBBS and DBS; (b) use the historical trip dataset to obtain the daily bike demands of each place in the past few days; (c) choose the maximum daily bike demands as the recommended initial bike amount of this place on this day. It assumes that the bike demands at each place on this day do not significantly exceed the maximum bike demands of this place in the past few days. This assumption has been proved to be correct, and only a few bike demands and trip demands are not met, which is described in Section 4.3.

The solution to future uncertainty with a maximum value is based on a consensus that "extremes occur rarely" (Hegerl et al., 2011; Broska et al., 2020). In this study, it means that "short-term futural states rarely exceed long-term historical thresholds". This consensus has been proved by the analysis based on bike sharing trip data and also has been indirectly reflected in Fig. 6. In addition, a basic logic is that "increasing bike amount helps or does not affect the satisfaction of travel demand, but will not harm the satisfaction of travel demand". Therefore, the bike amount in the short-term future can be determined according to the upper threshold of the long-term history. Our solution can effectively meet the travel demand and is feasible in actual operation.

The general framework can be established to minimize fleet size under the complete-information assumption and minimize fleet size under an incomplete-information scenario if a parameter of historical dataset period $u$ is added. Historical dataset period $u$ has two values because of the two types of minimum fleet size. The existing studies (Vazifeh et al., 2018; Gu et al., 2019;

Luo et al., 2020) only consider the complete-information assumption. However, since future demands cannot be completely predicted, these existing studies instead assume that the trip demands of the previous day will be repeated on this day. So the value of $u$ under the complete-information assumption is 1 day, which uses the trip demands of the previous day to determine the initial bike allocation of this day. The value of $u$ under the incomplete-information scenario is 7 days, which uses the trip demands of the previous 7 days to determine the initial bike allocation of this day. The value conditions are shown in Formula (2).

$$u = \begin{cases} 1 \text{ day, under the complete} - \text{information assumption} \\ 7 \text{ days, under an incomplete} - \text{information scenario} \end{cases} \quad (2)$$

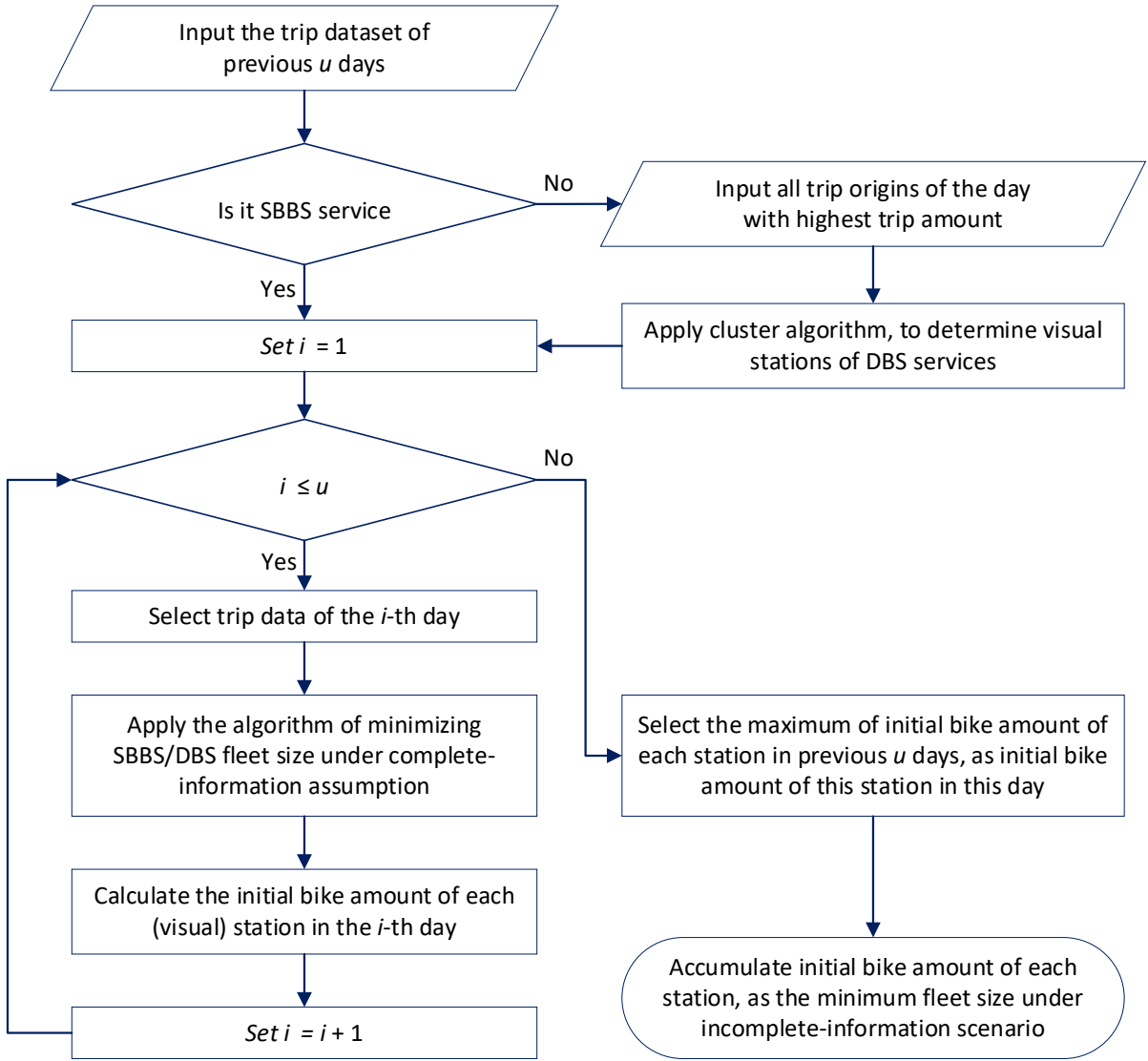

**Fig. 3.** Minimizing fleet size algorithm under the incomplete-information scenario (a general framework).

Besides, a general node or spatial object is also required for the bike allocation of both SBBS and DBS. In our study, a virtual station is proposed as a service node and spatial object, which can handle the uncertainty of the origin and destination locations of DBS trips. In SBBS, bikes are allocated at stations, while in DBS, bikes are allocated at virtual stations. A virtual station (Hua et al., 2020) is defined as is a set of bike positions, which is used as the spatial object of DBS. SBBS bikes are concentrated at the stations, and the stations are suitable spatial objects of SBBS bike allocation. The free-floating characteristic of DBS required that the city should be divided into thousands of small zones, which are the spatial objects of DBS bike allocation. Hua et al. (2020) have employed K-means clustering to recognize virtual stations for DBS parking analysis. However, there are some occasional trip demands in remote places, and these trips have great impacts on the clustering results of the virtual station. Therefore, an advanced clustering algorithm is proposed to identify virtual stations as DBS spatial objects.

The advanced clustering algorithm is K-means clustering with removing noise (K-means-RN) method, and the K-means-RN procedures of identifying DBS virtual stations are as follows. (a) Select the dataset of the day with the highest trip amount in the previous few days. (b) Apply Density-based spatial clustering of applications with noise (DBSCAN) clustering to all trip origins, with the parameters of radius 250 meters and minimum points 5. (c) Identify the occasional trip demands in the remote places as the DBSCAN noise points and remove these noise points. (d) Apply K-means clustering to the remaining origins, use the elbow method to get the appropriate cluster amount 4,000, and use cluster centers as the positions of the virtual stations. (f) Use the intersection of the Voronoi diagram and 250-meter buffer of each cluster center, to determine the service area of the 4,000 virtual stations.

### 3.3. Bike allocation and rebalancing

The large-scale determining algorithm for fleet allocation and bike rebalancing is shown in Fig. 4. Improving bike allocation would change the bike amounts in thousands of stations, which is essential for bike rebalancing. There are two types of rebalancing: static rebalancing at midnight and dynamic rebalancing in the daytime. This study focuses on static rebalancing, which determines the initial bike allocation at midnight. At midnight of this day, the company conducts the plan of initial bike allocation. After all users finish their trips, the bike distribution converts from initial bike allocation to final bike allocation at the end of this day. For meeting next-day demands, the company needs to reposition bikes from the final bike allocation of this day to the initial bike allocation of the next day.

It is worth mentioning that adding bikes to a place will not make trip demands unmet. Only when the predicted bike demands of the next day are not fully satisfied by the final bike allocation of this day, rebalancing trucks need to transfer bikes to fill the bike gap. For bike allocation and rebalancing, the spatial object of SBBS is the station, and the spatial object of DBS is the visual station. Bike sharing operators use trucks to visit SBBS stations or DBS virtual stations and load or unload bikes at these stations to meet the trip demands.

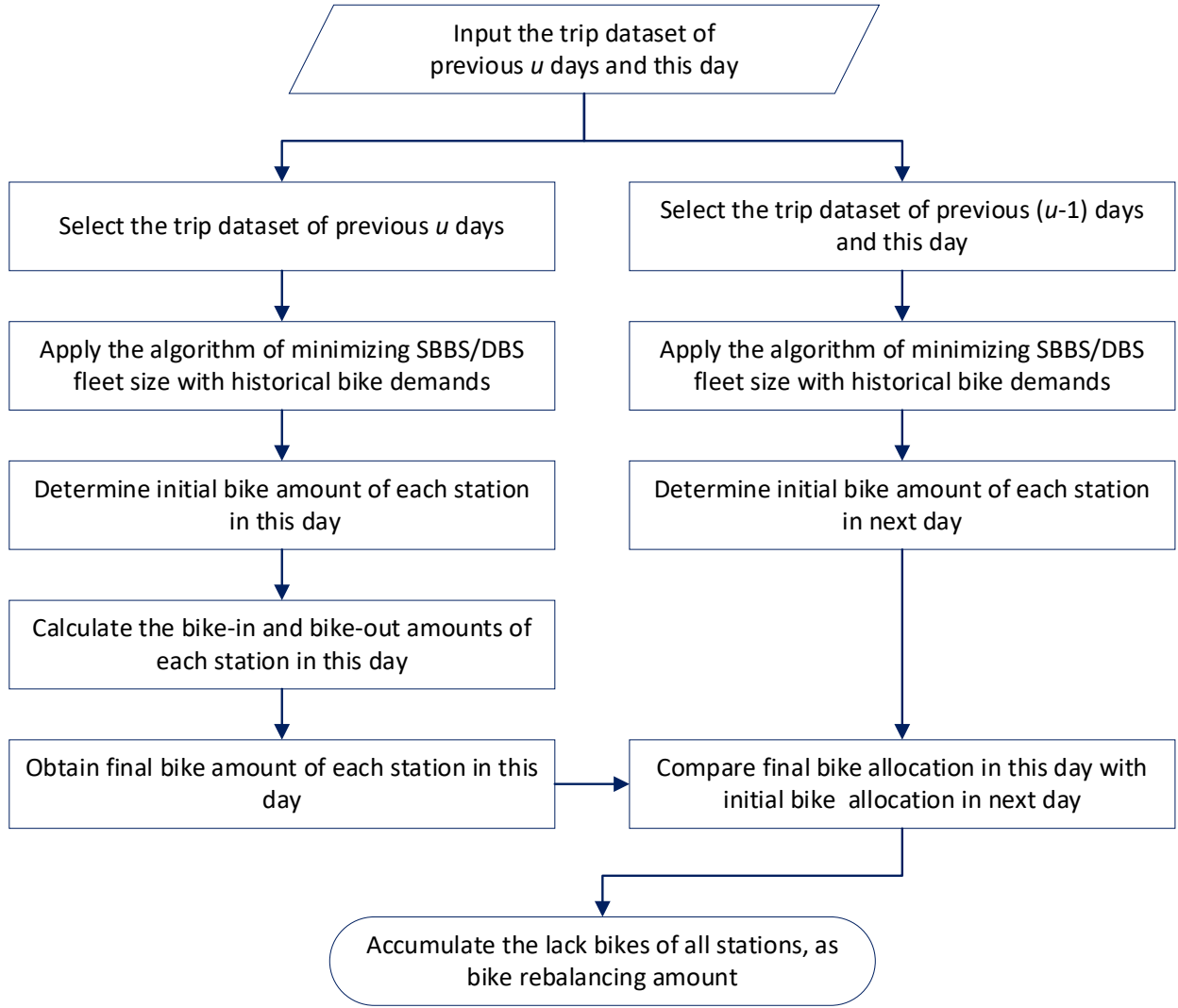

**Fig. 4.** The flowchart for determining bike allocation and rebalancing.

The proposed approach to minimize fleet size can solve large-scale problems and overcome future uncertainty. The existing fleet size studies rarely consider the large-scale problem, but many shared mobility companies provide large-scale services. The large-scale problem is a key challenge for shared mobility's operation management, and this paper addresses this challenge with a novel data-driven heuristic algorithm. Meanwhile, none of the existing minimum fleet size studies considers future uncertainty. All those current approaches would reject a large percentage of travel demand, which leads them to be completely infeasible in real-world company operation. Our proposed method solves the trade-off between reducing fleet size and meeting future travel demand and does not have the drawback of a high unmet trip ratio. Shared mobility companies such as bike sharing can directly apply our minimizing fleet size method as the daily initial inventory level for shared vehicle allocation.

## 4. Results

### 4.1. *Data and descriptive analysis*

This study possesses both SBBS and DBS trip data. Nanjing Public Bicycle Company Limited provided four years of SBBS trip data, from April 2016 to February 2020. The dataset has 143.0 million trip records. The fields of SBBS trip data include bike ID, user ID, start time, start station (origin), end time, and end station (destination). The company also provides the station information of all 1,447 stations, and the corresponding fields include station name, latitude, and longitude. In addition, Nanjing Transportation Bureau provided two weeks of DBS trip data, from January 5 to January 18 in 2020. The DBS dataset has 3.0 million trip records in the two weeks. The fields of DBS trip data include bike ID, company ID, start time, start longitude, start latitude, end time, end longitude, and end latitude. The personal information of DBS users was not disclosed. Three types of defective trips in these datasets are eliminated. (a) The start time or end time is missing. (b) The origin or destination is missing. (c) The origin or destination is not in Nanjing.

Bice sharing services have not only massive trips at the demand level but also large-scale fleets at the supply level. Nanjing has three DBS companies: Hellobike and Mobike entered Nanjing in 2017, and Didibike entered Nanjing in 2019. Based on company reports, Nanjing has an actual fleet size of 62,083 SBBS bikes and 350,642 DBS bikes. As for bike efficiency, each SBBS bike serves 5.9 trips per week, while each DBS bike serves 4.3 trips per week. A SBBS bike can serve more trips than a DBS bike, indicating that the SBBS bike fleet is in a more efficient state. For the given two-week data, it is computed that the active fleet size for SBBS and DBS are 31,105 and 204,638, respectively. Hence, only 50.1% of SBBS bikes and 58.4% of DBS bikes are active in this fortnight. In other words, 49.9% of SBBS bikes and 41.6% of DBS bikes are idle in the research period. This reflects the inefficient operation of bike sharing and the large number of idle bikes that should be reduced or activated.

Fig. 5 is plotted to facilitate understanding the relations between active fleet size and trip amount of SBBS and DBS. The actual fleet size fluctuates sharply together with the trip amount. There is a strong positive correlation between the trip amount and active fleet size. More trips result in more bike usage. But the volatility of trip amount is higher than that of active fleet size. In SBBS, the coefficient of variation (CV) of the trip amount is 0.307, much higher than the CV of active fleet size 0.219. In DBS, the coefficient of variation (CV) of the trip amount is 0.351, much higher than the CV of active fleet size 0.182. A key finding is that the future uncertainty of travel demand is much higher than that of bike supply. This inspires us that considering the bike-supply level rather than the travel-demand level could be a better option for dealing with future uncertainty.

Fig. 6 presents investigating the fluctuation and volatility of SBBS trip demands data, which suggests the effectiveness of our proposed method. First, it is noticed that there is a sharp drop in the trip amount in February, which is due to the holiday impact of the Chinese New Year. However, despite the daily fluctuation, the highest daily trip amount of each week is relatively stable. Regarding the daily variation, we can compute that the relative root mean square error of two continuous days is as large as 26.2%. In contrast, the relative root mean square error of these two continuous weeks is only 8.8%. The weekly volatility of trip demands is relatively acceptable in this study. Even though the monthly volatility of trip demands is potentially less, a month is so long that it is difficult to keep up to date with the temporal changes in trip demands. These findings indicate that applying historical data of the previous week to predict future trip demands is feasible and appropriate.

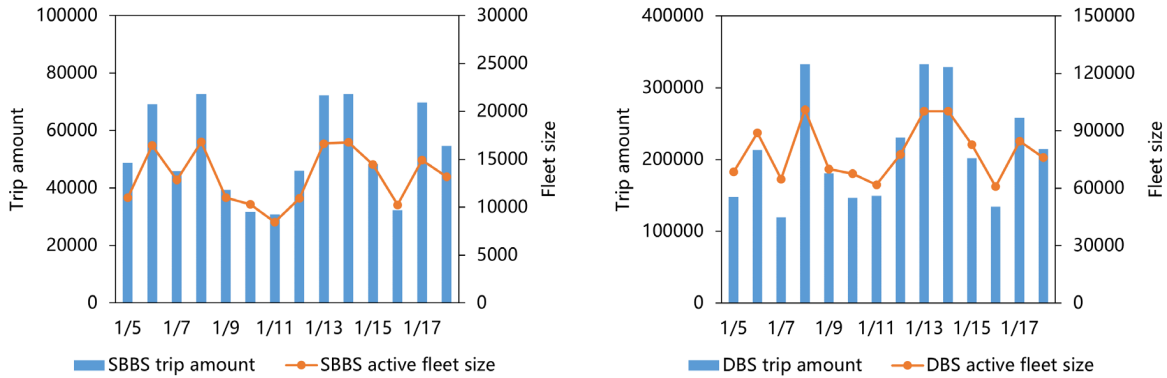

(a) Daily trips and used bikes of SBBS      (b) Daily trips and used bikes of DBS

**Fig. 5.** The trip demands and active fleet size in bike sharing during the fortnight in 2020.

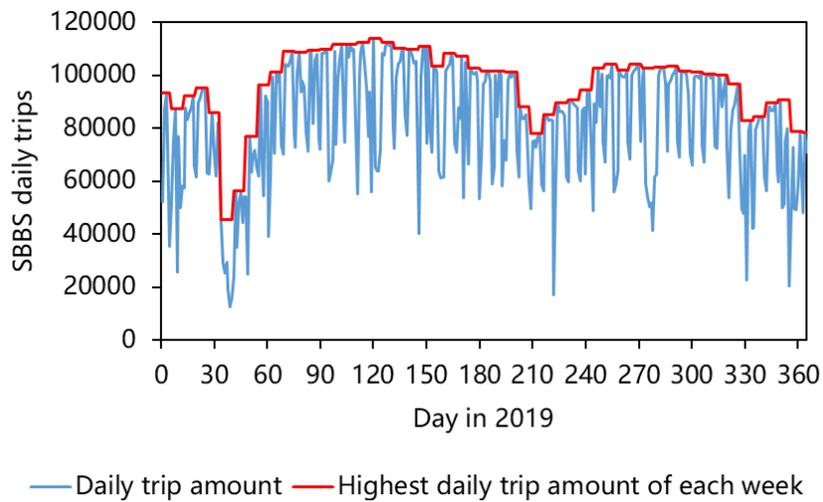

**Fig. 6.** The volatility of daily trip amount in Nanjing SBBS.

*4.2. Evaluation indicators*

The evaluation indicators for the bike allocation method are recommended fleet size, repositioning for the next day, and unmet trip ratio. Recommended fleet size and repositioning for the next day are easily calculated based on the large-scale bike allocation algorithm. The unmet trip ratio is the ratio of unmet trips to all trips of the actual trip dataset. The unmet trip ratio is the most important indicator of bike allocation results. In any initial bike allocation plan, there is always a part of future trip demands that will not be met. If the recommended bike allocation rejects too many user trip requests, the users would choose other mobility services and will no longer consider bike sharing. It is assumed that the unmet trip ratio should not be higher than 5%; otherwise, the result is unacceptable.

There are two different methods for calculating the unmet trip ratio for SBBS and DBS. In the case studies of Nanjing, there are 1,447 fixed stations for SBBS and 4,000 virtual stations for DBS. For SBBS, the station-based method is used for calculating this ratio, and all bikes are parked in stations. If there is an amount gap $g$ between bike demands and bike allocation in a station, the trips of the last $g$ bikes used in this station will not be met in the minimum fleet size algorithm under the complete-information assumption. For DBS, the bike-based method is used for calculating this ratio, and all bikes can be parked anywhere, including the centers of virtual stations. Whether the DBS initial bike allocation can meet the trip demands is determined by each DBS bike position, which is different from SBBS. A DBS bike in a virtual station may meet the trip demands generated in another virtual station. It reflects the free-floating mobility of DBS and the scattered nature of its trip demands.

*4.3. The results of two minimizing fleet size algorithms*

Minimizing fleet size algorithm under the complete-information assumption is applied to the bike sharing trip dataset in the fortnight of 2020, and the results are shown in Fig. 7. The minimum fleet size is found to be significantly smaller than the active fleet size, which means that this bike allocation method can greatly reduce bike supply without losing trip demands.

The higher the daily trip amount, the higher the minimum fleet size required to meet all trip demands, which is consistent with common sense. For SBBS, the Pearson correlation coefficient between daily trip amount and minimum fleet size is 0.95. For DBS, this correlation coefficient between daily trip amount and minimum fleet size is 0.98. Performing linear regression on the daily trip amount and minimum fleet size, the R-square in SBBS is 0.90, and the R-square in DBS is 0.95. The results of the correlation coefficient and the R-square imply that trip amount and minimum fleet size have a positive linear correlation.

Fleet reduction ratio, as an indicator of bike redundancy in the fleet, is the ratio of the difference between active fleet size and minimum fleet size to active fleet size. In SBBS, the fleet reduction ratio is 32.2%. In DBS, the fleet reduction ratio is 63.7%. Compared with SBBS, DBS can reduce bike fleet size to a greater extent, which indicates that more bikes of DBS are redundant.

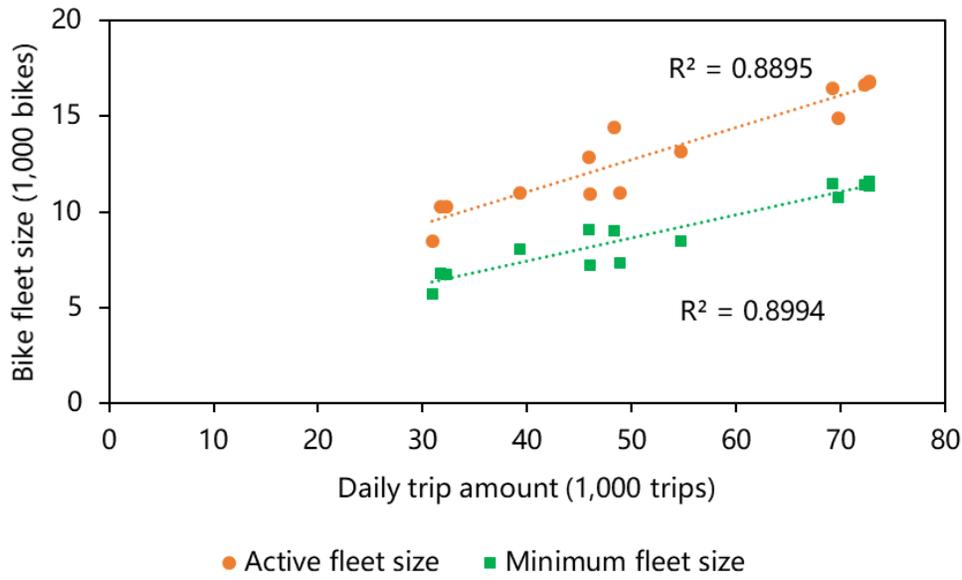

(a) The relationship of trip demands and fleet size in SBBS

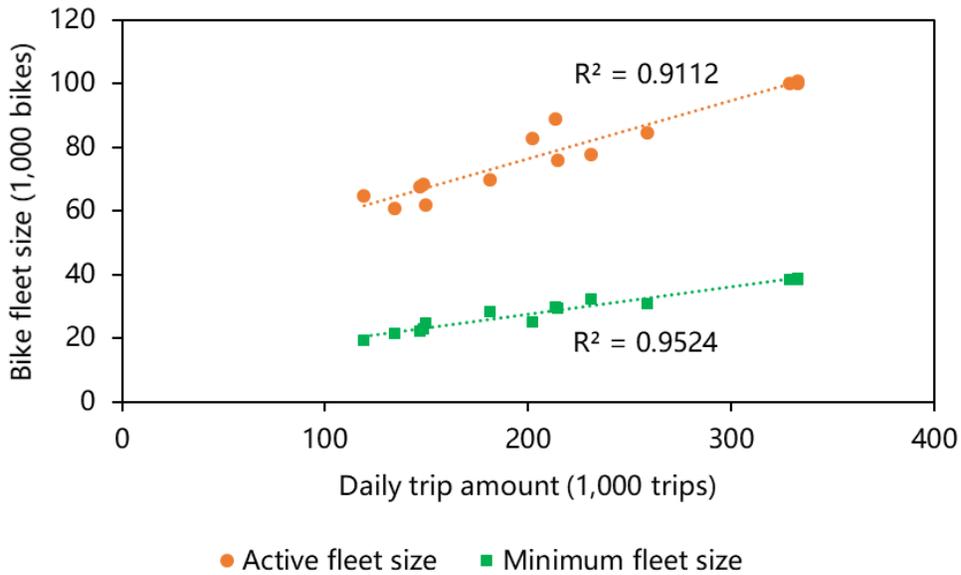

(b) The relationship of trip demands and fleet size in DBS

**Fig. 7.** The daily trip and minimum fleet size under the complete-information assumption.

    The difference in fleet reduction ratio of SBBS and DBS is due to the different mobility nature and bike efficiency of the two services. SBBS mobility has the restriction of the fixed station, and SBBS bikes can only match the trip demands of its station. But DBS bikes are free-floating and can match any trip demand within acceptable walking distance. SBBS bikes are more challenging to connect trips, and SBBS has a higher bike turnover than DBS (11.8 vs. 7.3 trips weekly per bike). So the redundancy of SBBS is lower. DBS has stronger connecting possibilities and lower

bike efficiency, so reducing a larger proportion of DBS fleet size can also meet all trip demands.

Table 1 shows the detailed results of minimizing fleet size under the complete-information assumption and minimizing fleet size under the incomplete-information scenario. Minimizing fleet size algorithm under the complete-information assumption is not practical in actual operation because the future demands cannot be completely predicted. So this fully-informed algorithm commonly assumes that this day's trips would repeat on the next day. In the actual application of this fully-informed algorithm, the trip dataset of this day is used to determine the initial bike allocation of the next day.

**TABLE 1** Result performance comparison of two minimizing fleet size algorithms.

(a) Two minimizing fleet size algorithms applied in SBBS

| Date | Minimizing fleet size under the complete-information assumption | | | Minimizing fleet size under the incomplete-information scenario | | |
|---|---|---|---|---|---|---|
| | Recommended fleet size (1,000 bikes) | Repositioning for next day (1,000 bikes) | Unmet trip ratio | Recommended fleet size (1,000 bikes) | Repositioning for next day (1,000 bikes) | Unmet trip ratio |
| 2020/1/12 | 5.7 | 5.2 | 22.5% | 16.6 | 3.1 | 1.1% |
| 2020/1/13 | 7.2 | 8.8 | 27.2% | 16.6 | 4.9 | 3.2% |
| 2020/1/14 | 11.4 | 6.2 | 13.7% | 16.7 | 5.1 | 3.0% |
| 2020/1/15 | 11.4 | 5.7 | 9.6% | 17.3 | 5.3 | 1.4% |
| 2020/1/16 | 9.0 | 3.6 | 11.8% | 16.7 | 2.9 | 0.8% |
| 2020/1/17 | 6.7 | 8.6 | 25.9% | 16.2 | 5.3 | 3.0% |
| 2020/1/18 | 10.7 | 4.5 | 13.3% | 17.1 | 3.8 | 1.6% |
| Average | **8.9** | **6.1** | **17.7%** | **16.8** | **4.3** | **2.0%** |

(b) Two minimizing fleet size algorithms applied in DBS

| Date | Minimizing fleet size under the complete-information assumption | | | Minimizing fleet size under the incomplete-information scenario | | |
|---|---|---|---|---|---|---|
| | Recommended fleet size (1,000 bikes) | Repositioning for next day (1,000 bikes) | Unmet trip ratio | Recommended fleet size (1,000 bikes) | Repositioning for next day (1,000 bikes) | Unmet trip ratio |
| 2020/1/12 | 23.8 | 21.2 | 16.0% | 47.6 | 25.0 | 4.0% |
| 2020/1/13 | 31.2 | 24.3 | 16.7% | 49.4 | 27.1 | 3.7% |
| 2020/1/14 | 37.7 | 22.5 | 9.4% | 51.3 | 28.3 | 3.4% |
| 2020/1/15 | 37.4 | 11.3 | 7.8% | 53.2 | 24.4 | 2.6% |
| 2020/1/16 | 24.4 | 11.3 | 9.1% | 51.5 | 23.9 | 2.6% |
| 2020/1/17 | 20.7 | 19.9 | 18.0% | 50.9 | 25.6 | 2.9% |
| 2020/1/18 | 29.7 | 16.3 | 11.6% | 51.4 | 25.9 | 3.4% |
| Average | **29.3** | **18.1** | **12.7%** | **50.8** | **25.7** | **3.2%** |

Compared with the minimum fleet size algorithm under the complete-information assumption, the minimum fleet size algorithm under the incomplete-information scenario requires more bike supply. There is not much difference between the two minimum fleet size algorithms in bike repositioning. Most importantly, under the complete-information assumption, the unmet trip ratio is more than 12% and unacceptable. However, the unmet trip ratio under the incomplete-information scenario is less than 4% and acceptable. In Nanjing, SBBS has 62.1 thousand bikes, and DBS has 350.6 thousand bikes. For the minimum fleet size algorithm under the incomplete-information scenario, SBBS service uses 16.8 thousand (27.0%) bikes to meet 98.0% of trip demands, and DBS service uses 50.8 thousand (14.5%) bikes to meet 96.8% of trip demands. A small part of fleet size and reasonable bike allocation can effectively meet future trip demands.

In this study, only a few trip demands are not met, which is unavoidable in actual operation. Existing studies (O'Mahony and Shmoys, 2015; Goh et al., 2019; Negahban, 2019) have found that the observed datasets are trip demands that have been met, and there are still some potential trip demands that are not met in actual operation. Because of the uncertainty of the future, the future trip demands cannot be completely predicted and fully met. The only solution is determining a reasonable bike allocation plan based on the historical dataset to meet most of the future trip demands.

## 5. Sensitivity analysis

In this section, we change the parameters and assumptions in the minimizing fleet size algorithm to explore the robustness of this method in different scenarios. Firstly, the bike sharing strategy for reducing the virus transmission risk during COVID-19 are discussed by changing the parameter of bike usage interval. Secondly, the effect of a single platform for reducing fleet size is estimated by comparing a single platform for collaborative services and multiple companies operating independently.

### 5.1. Normal operation vs. user distancing during COVID-19

In different scenarios, the fleet of bike sharing should have different operational states. This user distancing requirement is that there should be enough time intervals between the former and latter users using the same bike (bike usage interval). Bikes with too short bike usage intervals are considered unsafe and may be at risk of coronavirus transmission. This study suggests that when a user scans a bike's code, he or she will be notified of the end time of the last trip of this bike. If the time interval is shorter than the threshold value of bike usage interval $c$, users will be prohibited from using this bike and be recommended to use another suitable bike to achieve social distancing during the COVID-19 pandemic.

Bike usage interval $c$ is used to distinguish between normal operation and user distancing during the COVID-19 pandemic. In normal operation, $c$ is set as 0 to improve bike turnover and fleet efficiency. During the pandemic, $c$ can be set as greater than 0 (such as 1 hour or 6 hours), to meet the user distancing requirement. According to the surface stability study of SARS-CoV-2 coronavirus (van Doremalen et al., 2020), 6 hours are the half-life of this COVID-19 coronavirus. So 6 hours are selected as one recommended value of bike usage interval $c$ during COVID-19, and 1 hour is also selected as another *c* value for comparison.

Fig. 8 shows the results of the DBS fleet size with different bike usage intervals. When *c* is 1 hour, fleet size has increased by 17.6%. When *c* is 6 hours, fleet size has increased by 74.9%. Applying user distancing strategy in bike sharing requires more bike fleet to meet trip demands. However, when *c* is 6 hours, the fleet size is still smaller than the active fleet size and far smaller than the actual fleet size. The user distancing strategy is practical and feasible. This finding demonstrates that reasonable allocation and efficient repositioning of the existing bike fleet can use idle and redundant bikes to support the social distancing strategy during the COVID-19 pandemic.

User distancing can help activate many idle bikes and cooperate with other COVID-19 control strategies. In normal operation, about 40%-50% shared bikes in Nanjing are idle. During the severe period of COVID-19, travel demand would drop, and more bikes would become idle. User distancing strategy that lengthen bike usage interval could activate those idle bikes. So bike sharing companies can embrace user distancing strategy to reduce the virus transmission risk during the pandemic. In addition, bike disinfection is also an essential type of COVID-19 control strategies. But bike disinfection requires hiring many staff to drive disinfecting vehicles around the city to clean hundreds of thousands of bikes in thousands of places. Bike disinfection is expensive, and it is also difficult to disinfect all used bikes in a timely manner. Therefore, it is a possible solution to combine bike disinfection and user distancing, which not only activates massive idle bikes, but also sustainably reduces the virus transmission risk at low cost.

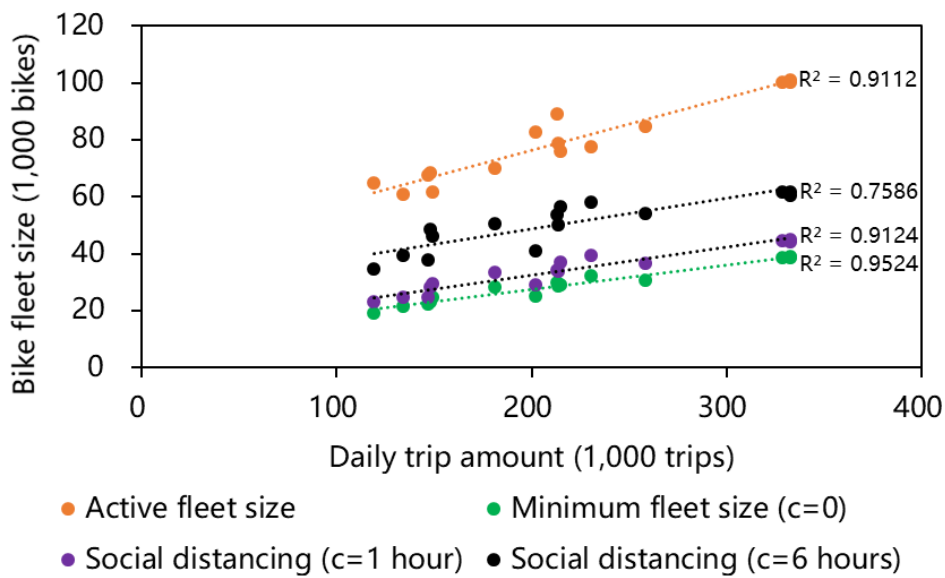

**Fig. 8.** The fleet size results with different bike usage intervals *c*.

### 5.2. Single platform vs. multiple companies

In the previous parts, a unified platform is used to meet the large-scale user demands of all companies. That is, a user can use the bikes of any company on this platform. Nanjing has three DBS companies, Hellobike, Mobike, and Didibike. Taking into account their respective business interests, each company is now operating independently and competing with each other. The independent operation of multiple companies has a great negative impact on minimizing fleet size. Compared with the independent operation of multiple companies, the minimum fleet size of the single platform can be reduced by 44.6%. A unified platform can not only provide users with more bike availability and more convenient services but also significantly reduce the fleet size that enterprises need to supply. After the fleet size is significantly reduced, the occupied urban space and operating resources will also be greatly reduced.

A single platform for multiple bike sharing companies has taken shape. For example, users can scan codes in Alipay (China's most popular mobile payment application) to unlock the bikes of multiple companies and pay for these trips. But the main consumption habit of DBS users is buying monthly memberships instead of paying for each trip. The monthly membership of a company cannot be shared among other companies. This is the biggest challenge for the unified platform to become a reality. But the single unified platform is still a wonderful vision and a promising trend for shared micro-mobility, including both bike sharing and scooter sharing.

DBS has economies of scale. That is, a larger fleet size corresponds to a lower cost per trip. As shown in Fig. 9(a), Hellobike has the largest fleet size and the highest bike turnover, followed by Mobike, and Didibike has the smallest fleet size and the lowest bike turnover. The larger the fleet size of a company, the higher bike turnover and the lower bike cost of each trip. In addition, the larger the fleet size, the more efficiency and less redundancy of its bike fleet. As shown in Fig. 9(b)-(d), the larger the fleet size, the fewer ratio of bikes that can be reduced. The fleet reduction ratios of Hellobike, Mobike, and Didibike are 31.0%, 58.6%, and 65.5%, respectively. The economies of scale of DBS may lead to an oligopoly in the market, and companies will fall into the vicious competition that unreasonably increases bike supply and produces many negative externalities.

It is worth mentioning that this paper does not involve the feasibility study of establishing a single platform. The establishment of a single platform requires cooperation between bike sharing companies and the government on the practical level, and also requires economic analysis such as game theory on the theoretical level. Our study only illustrates that the idea of a single platform has begun to take shape and discusses the beneficial impact of establishing a single platform on reducing the fleet size of bike sharing services.

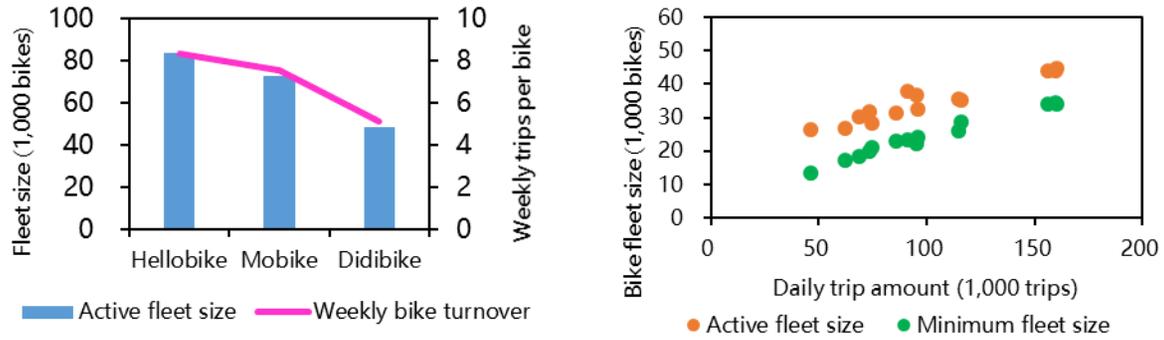

(a) Fleet size and bike turnover of 3 companies  (b) Minimum fleet size result of Hellobike

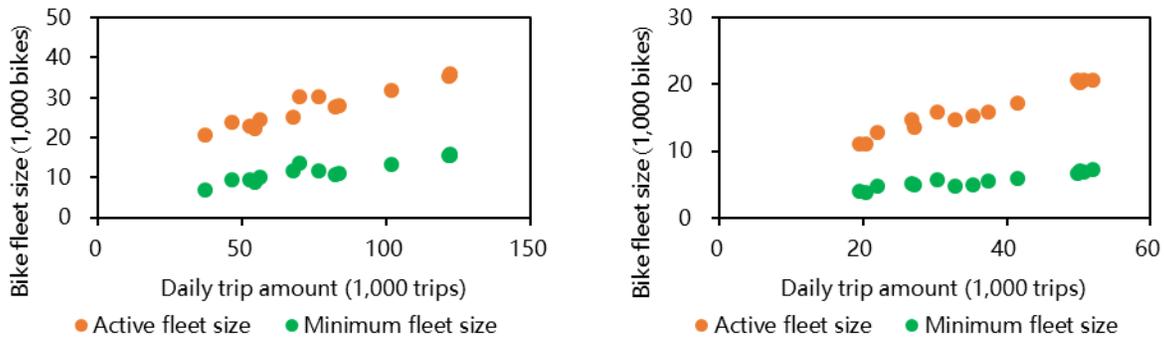

(c) Minimum fleet size result of Mobike  (d) Minimum fleet size result of Didibike

**Fig. 9.** The minimizing fleet size results in multiple companies.

## 6. Conclusion

Based on the trip datasets of DBS and SBBS in Nanjing, this study solves the large-scale minimum fleet size problem of bike sharing and provides the general framework of bike allocation. The minimizing fleet size algorithms under the complete-information assumption and the incomplete-information scenario are compared because of the volatility of trip demands. The key indicator of unmet trip ratio is used to validate the bike allocation method in meeting the uncertain future trip demands. The bike allocation method can be applied in the large-scale bike sharing service of millions of trips, thousands of stations, and hundreds of thousands of bikes. The proposed method can also be used for various scenarios, such as imposing social distancing policy under pandemic and establishing a single platform. The main findings are as follows.

The higher the trip amount, the larger the minimum fleet size to meet these demands. Because of the mobility nature and the difference in bike efficiency, DBS can reduce the bike fleet by greater than SBBS. The SBBS's fleet reduction ratio is 32.2%, while the DBS's fleet reduction ratio is 63.7%. However, future trip demands cannot be completely predicted and fully satisfied. The minimizing fleet size algorithm is improved by considering historical bike demands. Minimum fleet size and initial vehicle configuration are determined by the maximum bike demands in the previous week at each place. The minimum fleet size under an incomplete-information scenario is larger than that under the complete-information assumption, but the unmet trip ratio has sharply decreased from more than 12% to less than 4%. Minimizing fleet size algorithm under incomplete-information scenario can meet most of the future trip demands and is practical for company operation.

The COVID-19 pandemic has a huge impact on transportation, and how to avoid coronavirus transmission has become a critical problem for mobility services. By setting an appropriate value for bike usage interval, former and latter users of the same bike can maintain a sufficient time interval to achieve user distancing in bike sharing. The company will prohibit users from using unsafe bikes and recommend using nearby safe bikes. This study finds that user distancing strategy requires a larger fleet size of bike sharing, but the increased bike demands can be met by activating idle bikes. Although this study focuses on the social distancing strategy in bike sharing, it also has helpful inspirations for the epidemic response of other transportation modes such as scooter sharing, car sharing, taxi, and ride sourcing.

There are economies of scale in bike sharing. The more bikes, the higher the bike efficiency and the less fleet redundancy. There is a negative correlation between fleet size and fleet reduction ratio. The oligopoly feature and fierce competition among multiple enterprises are important reasons for the bike over-supply problem. To solve this problem, this study puts forward the idea of a unified platform to serve all users. In this single platform, all users can use any company's bikes to improve bike efficiency and reduce fleet size. After a single platform is built, the minimum fleet size that meets the same demands will drop by 44.6%.

Future uncertainty is the key to the minimizing fleet size problem, but our proposed method does not involve random variables or probability distributions. Random variables or probability distribution are the theoretical simulation of future travel demand, which is popular among the existing research but has questionable predicting results. Considering that future travel demand forecasting is a very difficult problem, the results of such a theoretical simulation are not very reliable. Besides, the idea of theoretical simulation requires four steps for vehicle allocation: "historical travel demand observation" - "theoretical simulation model establishing" – "future travel demand forecasting" - "future vehicle allocation planning", and these links are also not very reliable. On the contrary, our study is not concerned with the theoretical simulation of travel demand forecasting at the travel-demand level but focuses on evaluating vehicle allocation results at the bike-supply level. Because futural uncertainty of the bike-supply level is much less than that of the travel-demand level, we deal with future uncertainty as the difference in required bike amount between the former day and the latter day. The idea of our proposed method is simple and reliable, which only requires two steps for vehicle allocation: "historical travel demand observation" – "future vehicle allocation planning". Compared with the theoretical simulation idea such as random variables, our study has strong advantages and reliability because of directly matching supply and demand.

This research can be extended in the following aspects. Firstly, because of lacking the distribution data of bike parking spaces, the condition of bikes exceeding the parking limit is not considered. It is necessary to obtain comprehensive and accurate parking data, to improve the parking and repositioning of bike sharing. Secondly, the application of the unified platform in actual operation needs in-depth research. Due to economies of scale, the trip demands will increase after this single platform is built, and the existing fleet requires more reasonable bike allocation and repositioning. Moreover, how multiple companies appropriately distribute benefits in the platform is also a prerequisite for the platform's construction. Thirdly, this research proves that the greedy algorithm in SBBS can obtain the minimum fleet size, but the greedy algorithm in DBS may not definitely obtain the minimum fleet size. Other algorithms such as the Hopcroft–Karp algorithm can also be applied to the minimum fleet size problem of DBS, to compare with the results of the greedy algorithm. Finally, considering the similarities and differences in operational characteristics, this study on bike sharing can be extended to other shared mobility services with appropriate adaptations.

## Acknowledgments

The authors appreciate Nanjing Public Bicycle Company Limited and Nanjing Transportation Bureau for providing the data used in this study. This study is supported by the National Natural Science Foundation of China (52172316).

## References

Bao, J., Xu, C., Liu, P., Wang, W., 2017. Exploring Bikesharing Travel Patterns and Trip Purposes Using Smart Card Data and Online Point of Interests. Networks Spat. Econ. 17, 1231–1253. https://doi.org/10.1007/s11067-017-9366-x

Böcker, L., Anderson, E., Uteng, T.P., Throndsen, T., 2020. Bike sharing use in conjunction to public transport: Exploring spatiotemporal, age and gender dimensions in Oslo, Norway. Transp. Res. Part A Policy Pract. 138, 389–401. https://doi.org/10.1016/j.tra.2020.06.009

Böckin, D., Willskytt, S., André, H., Tillman, A.-M., Ljunggren Söderman, M., 2020. How product characteristics can guide measures for resource efficiency — A synthesis of assessment studies. Resour. Conserv. Recycl. 154, 104582. https://doi.org/https://doi.org/10.1016/j.resconrec.2019.104582

Broska, L.H., Poganietz, W.-R., Vögele, S., 2020. Extreme events defined—A conceptual discussion applying a complex systems approach. Futures 115, 102490. https://doi.org/https://doi.org/10.1016/j.futures.2019.102490

Bruck, B.P., Cruz, F., Iori, M., Subramanian, A., 2019. The static bike sharing rebalancing problem with forbidden temporary operations. Transp. Sci. 53, 882–896. https://doi.org/10.1287/trsc.2018.0859

Bucsky, P., 2020. Modal share changes due to COVID-19: The case of Budapest. Transp. Res. Interdiscip. Perspect. 8, 100141. https://doi.org/10.1016/j.trip.2020.100141

Caggiani, L., Camporeale, R., Marinelli, M., Ottomanelli, M., 2019. User satisfaction based model

for resource allocation in bike-sharing systems. Transp. Policy 80, 117–126. https://doi.org/10.1016/j.tranpol.2018.03.003

Caggiani, L., Colovic, A., Ottomanelli, M., 2020. An equality-based model for bike-sharing stations location in bicycle-public transport multimodal mobility. Transp. Res. Part A Policy Pract. 140, 251–265. https://doi.org/https://doi.org/10.1016/j.tra.2020.08.015

Chen, J., Zhou, D., Zhao, Y., Wu, B., Wu, T., 2020. Life cycle carbon dioxide emissions of bike sharing in China: Production, operation, and recycling. Resour. Conserv. Recycl. 162, 105011. https://doi.org/https://doi.org/10.1016/j.resconrec.2020.105011

Chen, W., Liu, Q., Zhang, C., Mi, Z., Zhu, D., Liu, G., 2020. Characterizing the stocks, flows, and carbon impact of dockless sharing bikes in China. Resour. Conserv. Recycl. 162, 105038. https://doi.org/https://doi.org/10.1016/j.resconrec.2020.105038

Choudhury, S.R., 2018. Still "a lot of space" for competition in China's bike-sharing frenzy, says start-up exec [WWW Document]. CNBC.

Datner, S., Raviv, T., Tzur, M., Chemla, D., 2019. Setting inventory levels in a bike sharing network. Transp. Sci. 53, 62–76. https://doi.org/10.1287/trsc.2017.0790

Dlugosch, O., Brandt, T., Neumann, D., 2020. Combining analytics and simulation methods to assess the impact of shared, autonomous electric vehicles on sustainable urban mobility. Inf. Manag. 103285. https://doi.org/https://doi.org/10.1016/j.im.2020.103285

Du, M., Cheng, L., Li, X., Tang, F., 2020. Static rebalancing optimization with considering the collection of malfunctioning bikes in free-floating bike sharing system. Transp. Res. Part E Logist. Transp. Rev. 141, 102012. https://doi.org/https://doi.org/10.1016/j.tre.2020.102012

Eren, E., Katanalp, B.Y., 2022. Fuzzy-based GIS approach with new MCDM method for bike-sharing station site selection according to land-use types. Sustain. Cities Soc. 76, 103434. https://doi.org/https://doi.org/10.1016/j.scs.2021.103434

Faghih-Imani, A., Eluru, N., 2016. Incorporating the impact of spatio-temporal interactions on bicycle sharing system demand: A case study of New York CitiBike system. J. Transp. Geogr. 54, 218–227. https://doi.org/10.1016/j.jtrangeo.2016.06.008

Fishman, E., Washington, S., Haworth, N., Watson, A., 2015. Factors influencing bike share membership: An analysis of Melbourne and Brisbane. Transp. Res. Part A Policy Pract. 71, 17–30. https://doi.org/10.1016/j.tra.2014.10.021

Fu, C., Zhu, N., Ma, S., Liu, R., 2022. A two-stage robust approach to integrated station location and rebalancing vehicle service design in bike-sharing systems. Eur. J. Oper. Res. 298, 915–938. https://doi.org/https://doi.org/10.1016/j.ejor.2021.06.014

Fuller, D., Gauvin, L., Kestens, Y., Daniel, M., Fournier, M., Morency, P., Drouin, L., 2011. Use of a New Public Bicycle Share Program in Montreal, Canada. Am. J. Prev. Med. 41, 80–83. https://doi.org/https://doi.org/10.1016/j.amepre.2011.03.002

Goh, C.Y., Yan, C., Jaillet, P., 2019. A Locational Demand Model for Bike-Sharing. SSRN Electron. J. https://doi.org/10.2139/ssrn.3311371

Gómez Márquez, H.R., López Bracho, R., Ramirez-Nafarrate, A., 2021. A simulation-
24

optimization study of the inventory of a bike-sharing system: The case of Mexico City Ecobici's system. Case Stud. Transp. Policy 9, 1059–1072. https://doi.org/https://doi.org/10.1016/j.cstp.2021.01.014

Gu, Z., Zhu, Y., Zhang, Y., Zhou, W., Chen, Y., 2019. Heuristic bike optimization algorithm to improve usage efficiency of the station-free bike sharing system in Shenzhen, China. ISPRS Int. J. Geo-Information 8, 239. https://doi.org/10.3390/ijgi8050239

He, M., Ma, X., Jin, Y., 2021. Station Importance Evaluation in Dynamic Bike-Sharing Rebalancing Optimization Using an Entropy-Based TOPSIS Approach. IEEE Access 9, 38119–38131. https://doi.org/10.1109/ACCESS.2021.3063881

Hegerl, G.C., Hanlon, H., Beierkuhnlein, C., 2011. Elusive extremes. Nat. Geosci. 4, 142–143. https://doi.org/10.1038/ngeo1090

Hu, R., Zhang, Z., Ma, X., Jin, Y., 2021. Dynamic Rebalancing Optimization for Bike-Sharing System Using Priority-Based MOEA/D Algorithm. IEEE Access 9, 27067–27084. https://doi.org/10.1109/ACCESS.2021.3058013

Hu, Y., Zhang, Y., Lamb, D., Zhang, M., Jia, P., 2019. Examining and optimizing the BCycle bike-sharing system – A pilot study in Colorado, US. Appl. Energy 247, 1–12. https://doi.org/https://doi.org/10.1016/j.apenergy.2019.04.007

Hua, M., Chen, X., Cheng, L., Chen, J., 2021. Should bike-sharing continue operating during the COVID-19 pandemic? Empirical findings from Nanjing, China. J. Transp. Heal. 23, 101264. https://doi.org/https://doi.org/10.1016/j.jth.2021.101264

Hua, M., Chen, X., Zheng, S., Cheng, L., Chen, J., 2020. Estimating the parking demand of free-floating bike sharing: A journey-data-based study of Nanjing, China. J. Clean. Prod. 244, 118764. https://doi.org/10.1016/j.jclepro.2019.118764

Jie, T., Wei, W., Jiang, L., 2020. A sustainability-oriented optimal allocation strategy of sharing bicycles: Evidence from ofo usage in Shanghai. Resour. Conserv. Recycl. 153, 104510. https://doi.org/https://doi.org/10.1016/j.resconrec.2019.104510

Kou, Z., Wang, X., Chiu, S.F. (Anthony), Cai, H., 2020. Quantifying greenhouse gas emissions reduction from bike share systems: a model considering real-world trips and transportation mode choice patterns. Resour. Conserv. Recycl. 153, 104534. https://doi.org/10.1016/j.resconrec.2019.104534

Kroes, J.R., Manikas, A.S., Gattiker, T.F., 2020. Generating efficient rebalancing routes for bikeshare programs using a genetic algorithm. J. Clean. Prod. 244, 118880. https://doi.org/https://doi.org/10.1016/j.jclepro.2019.118880

Kumar Dey, B., Anowar, S., Eluru, N., 2021. A framework for estimating bikeshare origin destination flows using a multiple discrete continuous system. Transp. Res. Part A Policy Pract. 144, 119–133. https://doi.org/https://doi.org/10.1016/j.tra.2020.12.014

Legros, B., 2019. Dynamic repositioning strategy in a bike-sharing system; how to prioritize and how to rebalance a bike station. Eur. J. Oper. Res. 272, 740–753. https://doi.org/10.1016/j.ejor.2018.06.051

Loidl, M., Witzmann-Müller, U., Zagel, B., 2019. A spatial framework for Planning station-based bike sharing systems. Eur. Transp. Res. Rev. 11, 9. https://doi.org/10.1186/s12544-019-0347-7

Luo, H., Kou, Z., Zhao, F., Cai, H., 2019. Comparative life cycle assessment of station-based and dock-less bike sharing systems. Resour. Conserv. Recycl. 146, 180–189. https://doi.org/https://doi.org/10.1016/j.resconrec.2019.03.003

Luo, H., Zhao, F., Chen, W.-Q., Cai, H., 2020. Optimizing bike sharing systems from the life cycle greenhouse gas emissions perspective. Transp. Res. Part C Emerg. Technol. 117, 102705. https://doi.org/https://doi.org/10.1016/j.trc.2020.102705

Lv, H., Wu, F., Luo, T., Gao, X., Chen, G., 2020. Hardness of and approximate mechanism design for the bike rebalancing problem. Theor. Comput. Sci. 803, 105–115. https://doi.org/https://doi.org/10.1016/j.tcs.2019.07.030

Ma, G., Zhang, B., Shang, C., Shen, Q., 2021. Rebalancing stochastic demands for bike-sharing networks with multi-scenario characteristics. Inf. Sci. (Ny). 554, 177–197. https://doi.org/https://doi.org/10.1016/j.ins.2020.12.044

Maleki Vishkaei, B., Mahdavi, I., Mahdavi-Amiri, N., Khorram, E., 2020. Balancing public bicycle sharing system using inventory critical levels in queuing network. Comput. Ind. Eng. 141, 106277. https://doi.org/https://doi.org/10.1016/j.cie.2020.106277

Masson, R., Trentini, A., Lehuédé, F., Malhéné, N., Péton, O., Tlahig, H., 2017. Optimization of a city logistics transportation system with mixed passengers and goods. EURO J. Transp. Logist. 6, 81–109. https://doi.org/https://doi.org/10.1007/s13676-015-0085-5

Negahban, A., 2019. Simulation-based estimation of the real demand in bike-sharing systems in the presence of censoring. Eur. J. Oper. Res. 277, 317–332. https://doi.org/10.1016/j.ejor.2019.02.013

O'Mahony, E., Shmoys, D.B., 2015. Data analysis and optimization for (Citi)bike sharing, in: Proceedings of the National Conference on Artificial Intelligence. pp. 687–694.

Otero, I., Nieuwenhuijsen, M.J., Rojas-Rueda, D., 2018. Health impacts of bike sharing systems in Europe. Environ. Int. 115, 387–394. https://doi.org/10.1016/j.envint.2018.04.014

Pal, A., Zhang, Y., 2017. Free-floating bike sharing: Solving real-life large-scale static rebalancing problems. Transp. Res. Part C Emerg. Technol. 80, 92–116. https://doi.org/10.1016/j.trc.2017.03.016

Ren, Yaping, Meng, L., Zhao, F., Zhang, C., Guo, H., Tian, Y., Tong, W., Sutherland, J.W., 2020. An improved general variable neighborhood search for a static bike-sharing rebalancing problem considering the depot inventory. Expert Syst. Appl. 160, 113752. https://doi.org/https://doi.org/10.1016/j.eswa.2020.113752

Ren, Y, Zhao, F., Jin, H., Jiao, Z., Meng, L., Zhang, C., Sutherland, J.W., 2020. Rebalancing Bike Sharing Systems for Minimizing Depot Inventory and Traveling Costs. IEEE Trans. Intell. Transp. Syst. 21, 3871–3882. https://doi.org/10.1109/TITS.2019.2935509

Rong, K., Xiao, F., Wang, Y., 2019. Redundancy in the sharing economy. Resour. Conserv. Recycl.

151, 104455. https://doi.org/https://doi.org/10.1016/j.resconrec.2019.104455

Simonetto, A., Monteil, J., Gambella, C., 2019. Real-time city-scale ridesharing via linear assignment problems. Transp. Res. Part C Emerg. Technol. 101, 208–232. https://doi.org/https://doi.org/10.1016/j.trc.2019.01.019

Soriguera, F., Jiménez-Meroño, E., 2020. A continuous approximation model for the optimal design of public bike-sharing systems. Sustain. Cities Soc. 52, 101826. https://doi.org/10.1016/j.scs.2019.101826

Swaszek, R.M.A., Cassandras, C.G., 2020. Receding Horizon Control for Station Inventory Management in a Bike-Sharing System. IEEE Trans. Autom. Sci. Eng. 17, 407–417. https://doi.org/10.1109/TASE.2019.2928454

Tao, J., Zhou, Z., 2021. Evaluation of Potential Contribution of Dockless Bike-sharing Service to Sustainable and Efficient Urban Mobility in China. Sustain. Prod. Consum. 27, 921–932. https://doi.org/https://doi.org/10.1016/j.spc.2021.02.008

Taylor, A., 2018. Bike Share Oversupply in China: Huge Piles of Abandoned and Broken Bicycles [WWW Document]. Atl.

Teixeira, J.F., Lopes, M., 2020. The link between bike sharing and subway use during the COVID-19 pandemic: The case-study of New York's Citi Bike. Transp. Res. Interdiscip. Perspect. 6, 100166. https://doi.org/10.1016/j.trip.2020.100166

Tian, Z., Zhou, J., Szeto, W.Y., Tian, L., Zhang, W., 2020. The rebalancing of bike-sharing system under flow-type task window. Transp. Res. Part C Emerg. Technol. 112, 1–27. https://doi.org/10.1016/j.trc.2020.01.015

van Doremalen, N., Bushmaker, T., Morris, D.H., Holbrook, M.G., Gamble, A., Williamson, B.N., Tamin, A., Harcourt, J.L., Thornburg, N.J., Gerber, S.I., Lloyd-Smith, J.O., de Wit, E., Munster, V.J., 2020. Aerosol and Surface Stability of SARS-CoV-2 as Compared with SARS-CoV-1. N. Engl. J. Med. 382, 1564–1567. https://doi.org/10.1056/NEJMc2004973

Vazifeh, M.M., Santi, P., Resta, G., Strogatz, S.H., Ratti, C., 2018. Addressing the minimum fleet problem in on-demand urban mobility. Nature 557, 534–538. https://doi.org/10.1038/s41586-018-0095-1

Wang, M., Zhou, X., 2017. Bike-sharing systems and congestion: Evidence from US cities. J. Transp. Geogr. 65, 147–154. https://doi.org/https://doi.org/10.1016/j.jtrangeo.2017.10.022

WIRED, 2017. Chinese Bike-Sharing Startup Mobike Has Its Eye on Expansion [WWW Document].

Xu, M., Meng, Q., Liu, Z., 2018. Electric vehicle fleet size and trip pricing for one-way carsharing services considering vehicle relocation and personnel assignment. Transp. Res. Part B Methodol. 111, 60–82. https://doi.org/10.1016/j.trb.2018.03.001

Yang, T., Yang, H., Wong, S.C., Sze, N.N., 2014. Returns to scale in the production of taxi services: An empirical analysis. Transp. A Transp. Sci. 10, 775–790. https://doi.org/10.1080/23249935.2013.794174

Yuan, M., Zhang, Q., Wang, B., Liang, Y., Zhang, H., 2019. A mixed integer linear programming

model for optimal planning of bicycle sharing systems: A case study in Beijing. Sustain. Cities Soc. 47, 101515. https://doi.org/https://doi.org/10.1016/j.scs.2019.101515

Zhang, J., Meng, M., 2019. Bike allocation strategies in a competitive dockless bike sharing market. J. Clean. Prod. 233, 869–879. https://doi.org/https://doi.org/10.1016/j.jclepro.2019.06.070

**Appendix**

The bike amount of our proposed method is equal to the minimum fleet size, whose derivation process is shown below. And our proposed method discussed here is minimizing fleet size algorithm under the complete-information assumption for SBBS.

In this bike allocation problem, there are two sets, the trip set and the bike set. The fleet assignment for satisfying travel demand is to use bikes to connect trips. The connection rule is that for the same bike, the destination of the previous trip and the origin of the next trip are the same, and the end time of the previous trip is not later than the start time of the next trip.

There must be a minimum fleet size that can satisfy all travel demand. The bike-connect-trip result of minimum fleet size can be described as using the fewest bikes to connect all trips. And our proposed method can be described as greedily deciding how bikes connect trips in order of time and bike.

The derivation process still requires the following three concepts.

(a) Greedy connection, it is defined as that a bike connects the next trip whose origin is the destination of the previous trip and whose start time is not earlier than and closest to the end time of the previous trip.

(b) Spatio-temporal intersection, it means that two bikes are parked together in a certain place at a certain moment. The spatio-temporal position of these two bikes are intersected. As shown in Appendix figure 1(a), Bike 1 and Bike 2 are both parked in Station 3 between the end time of Trip 2 and the start time of Trip 3.

(c) Break-reconnection, it means that the later trips of two bikes with spatio-temporal intersection can be exchanged. As shown in Appendix figure 1, Trip 3 and Trip 4 are exchanged for Bike 1 and Bike 2.

Break-reconnection would not change the bike amount, nor will it affect the satisfaction of travel demand. This is obvious because two bikes at the same time and the same place are equivalent in terms of bike-connect-trip. This equivalence is shown very intuitively in Appendix figure 1. In connection result A, Bike 1 connects Trip 1 and Trip 4, and Bike 2 connects Trip 2 and Trip 3. In connection result B, Bike 1 connects Trip 1 and Trip 3, and Bike 2 connects Trip 2 and Trip 4. The connection result A is not a greedy connection for Bike 1, and the connection result B is a greedy connection for Bike 1. Break-reconnection changes the connection result A into the connection result B, which does not affect bike amount and travel demand satisfaction.

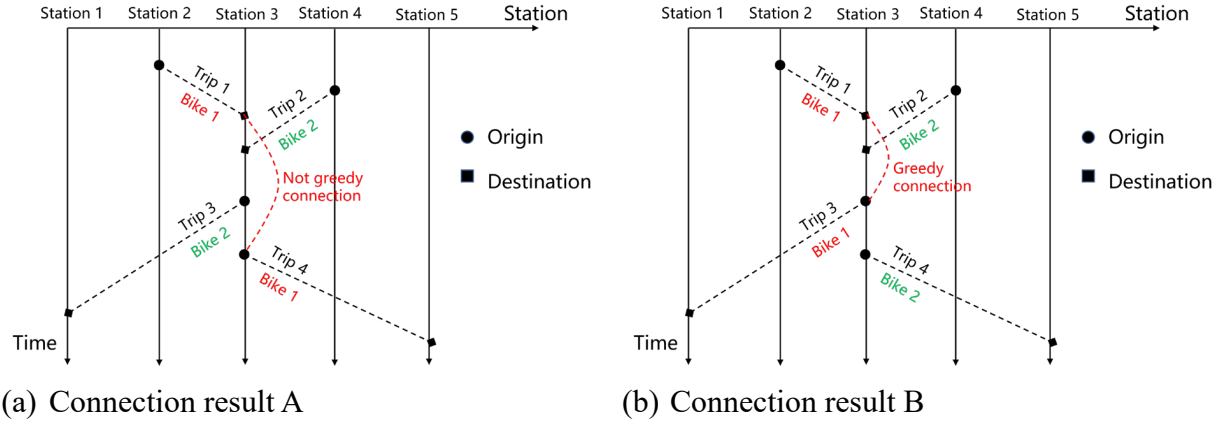

(a) Connection result A           (b) Connection result B

Appendix figure 1. Schematic of break-reconnection.

    We can show that, based on the above concepts, the bike-connect-trip results of minimum fleet size can be transformed into the bike-connect-trip results of our proposed method. First, it is assumed that the bike allocation is already in the state of minimum fleet size, which means that the minimum number of bikes have connected all trips. Then, the following process steps based on the principle of break-reconnection and iteration can be performed to change the connection results of the minimum fleet size into the connection results of our proposed method.

    (a) The first trip at the time level is selected, and it is the previous trip of the first bike. (b) The first bike is conducted with break-reconnection so to get the greedy connection at the destination of the previous trip, and the bike is moved to the destination of the new latter trip. (c) Step (b) is repeated until the new connection of the first bike cannot be made, at which time the connected trips are all trips connected by the first bike in our proposed method. (d) All connected trips are removed from the trip set. (e) Steps (a)(b)(c)(d) are repeated until the trip set is empty. After these process steps, the bike amount remains the same as the minimum fleet size, but the bike-connect-trip results of minimum fleet size have been transformed into the results of our proposed method.